\documentstyle[]{mn}
\newif\ifAMStwofonts

\def\etal{{\rm et al.}}

\def\simgt{\mathrel{\spose{\lower 3pt\hbox{$\sim$}}
        \raise 2.0pt\hbox{$>$}}}
\def\simlt{\mathrel{\spose{\lower 3pt\hbox{$\sim$}}
        \raise 2.0pt\hbox{$<$}}}

\ifoldfss
  \newcommand{\rmn}[1] {{\rm #1}}

  \ifCUPmtlplainloaded \else
    \NewTextAlphabet{textbfit} {cmbxti10} {}
    \NewTextAlphabet{textbfss} {cmssbx10} {}
    \NewMathAlphabet{mathbfit} {cmbxti10} {} 
    \NewMathAlphabet{mathbfss} {cmssbx10} {} 
  \fi
  \ifAMStwofonts
    \ifCUPmtlplainloaded \else
      \NewSymbolFont{upmath} {eurm10}
      \NewSymbolFont{AMSa} {msam10}
      \NewMathSymbol{\upi}     {0}{upmath}{19}
      \NewMathSymbol{\umu}     {0}{upmath}{16}
      \NewMathSymbol{\upartial}{0}{upmath}{40}
      \NewMathSymbol{\leqslant}{3}{AMSa}{36}
      \NewMathSymbol{\geqslant}{3}{AMSa}{3E}

      \let\leq=\leqslant 
       
    \fi
  \fi
\fi 

\ifnfssone
  \newmathalphabet{\mathit}
  \addtoversion{normal}{\mathit}{cmr}{m}{it}
  \addtoversion{bold}{\mathit}{cmr}{bx}{it}
  \newcommand{\rmn}[1] {\mathrm{#1}}

  \newmathalphabet{\mathbfit} 
  \addtoversion{normal}{\mathbfit}{cmr}{bx}{it}
  \addtoversion{bold}{\mathbfit}{cmr}{bx}{it}
  \newmathalphabet{\mathbfss} 
  \addtoversion{normal}{\mathbfss}{cmss}{bx}{n}
  \addtoversion{bold}{\mathbfss}{cmss}{bx}{n}
  \ifAMStwofonts
    \ifCUPmtlplainloaded \else
      \UseAMStwoboldmath
      \makeatletter
      \new@mathgroup\upmath@group
      \define@mathgroup\mv@normal\upmath@group{eur}{m}{n}
      \define@mathgroup\mv@bold\upmath@group{eur}{b}{n}
      \edef\UPM{\hexnumber\upmath@group}
      \new@mathgroup\amsa@group
      \define@mathgroup\mv@normal\amsa@group{msa}{m}{n}
      \define@mathgroup\mv@bold\amsa@group{msa}{m}{n}
      \edef\AMSa{\hexnumber\amsa@group}
      \makeatother
      \mathchardef\upi="0\UPM19
      \mathchardef\umu="0\UPM16
      \mathchardef\upartial="0\UPM40
      \mathchardef\leqslant="3\AMSa36
      \mathchardef\geqslant="3\AMSa3E

      \let\leq=\leqslant 

    \fi
  \fi
\fi 

\ifnfsstwo
  \newcommand{\rmn}[1] {\mathrm{#1}}

  \DeclareMathAlphabet{\mathbfit}{OT1}{cmr}{bx}{it}
  \SetMathAlphabet\mathbfit{bold}{OT1}{cmr}{bx}{it}
  \DeclareMathAlphabet{\mathbfss}{OT1}{cmss}{bx}{n}
  \SetMathAlphabet\mathbfss{bold}{OT1}{cmss}{bx}{n}
  \ifAMStwofonts
    \ifCUPmtlplainloaded \else
      \DeclareSymbolFont{UPM}{U}{eur}{m}{n}
      \SetSymbolFont{UPM}{bold}{U}{eur}{b}{n}
      \DeclareSymbolFont{AMSa}{U}{msa}{m}{n}
      \DeclareMathSymbol{\upi}{0}{UPM}{"19}
      \DeclareMathSymbol{\umu}{0}{UPM}{"16}
      \DeclareMathSymbol{\upartial}{0}{UPM}{"40}
      \DeclareMathSymbol{\leqslant}{3}{AMSa}{"36}
      \DeclareMathSymbol{\geqslant}{3}{AMSa}{"3E}

      \let\leq=\leqslant 

    \fi
  \fi
\fi 

\ifCUPmtlplainloaded \else
  \ifAMStwofonts \else 
    \def\upi{\pi}
    \def\umu{\mu}
    \def\upartial{\partial}
  \fi
\fi

\title[A determination of continuum source size in Q2237+0305]
  {A gravitational microlensing determination of continuum source size in Q2237+0305}
\author[J. S. B. Wyithe et al.]
  {J.~S.~B.~Wyithe$^1$, 
  R.~L.~Webster$^1$,
  E.~L.~Turner$^2$,
  D.~J.~Mortlock$^1$	\\
  $^1$ School of Physics, The University of Melbourne, Parkville, Vic, 3052, 
Australia\\
  $^2$ Princeton University Observatory, Peyton Hall, Princeton, NJ 08544, USA\\ 
 Email: swyithe@physics.unimelb.edu.au, rwebster@physics.unimelb.edu.au, elt@astro.princeton.edu, dmortloc@physics.unimelb.edu.au }
\date{Accepted Received}
\pagerange{\pageref{firstpage}--\pageref{lastpage}}
\pubyear{1999}

\def\LaTeX{L\kern-.36em\raise.3ex\hbox{a}\kern-.15em
    T\kern-.1667em\lower.7ex\hbox{E}\kern-.125emX}

\begin{document}

\label{firstpage}

\maketitle

\begin{abstract}
Following the detection of a gravitational microlensing high magnification event (HME) in Q2237+0305A attempts have been made to place limits on the dimensions of the quasar continuum source. The analyses have studied either the observed event magnitude or the event duration. The latter approach has been hampered by lack of knowledge about the transverse velocity of the lensing galaxy. We obtain both upper and lower statistical limits on the size of the continuum source from the observed HME using determinations of transverse velocity obtained from the published monitoring data. Our calculations take account of the caustic orientation as well as the component of the caustic velocity that results from stellar proper motions. Our determination of source size relies on an estimated duration of 52 days for the HME, and so will be refined when more HMEs are observed. We find that the upper and lower limits on the magnified region of the R-band continuum source are $6\times 10^{15}$ and $2\times 10^{13}\,cm$ respectively (99\% confidence). Through consideration of the joint probability for source size and mean microlens mass we find that the mean mass lies between  $\sim 0.01M_{\odot}$ and $\sim 1M_{\odot}$ (95\% confidence).   

\end{abstract}


\section{Introduction}

The object Q2237+0305 comprises a source quasar at redshift $z=1.695$ that is gravitationally lensed by a foreground galaxy ($z=0.0394$) producing 4 images with separations of $\sim 1''$.  Each of the 4 images are observed through the bulge of the galaxy, which has an optical depth in stars that is of order unity (eg. Kent \& Falco 1988; Schneider et al. 1988; Schmidt, Webster \& Lewis 1998). In addition, the proximity of the lensing galaxy means that the projected transverse velocity at the source is an order of magnitude higher than for a typical lens configuration. The combination of these considerations make Q2237+0305 the ideal object from which to study microlensing. Indeed, Q2237+0305 is the only object in which cosmological microlensing has been confirmed (Irwin et al. 1989; Corrigan et al. 1991). In 1988 a high magnification event (HME) was observed in image A. It had a measured rise time of $\sim 26$ days, and was followed by an event having a decline time of $\la 3$ months. The resulting light-curve has been interpreted as a double peaked event corresponding to the source having crossed two fold caustics inside a cusp (Witt \& Mao 1994), implying that the peak was caused by a single caustic crossing. The observation thus provides an estimate of the event crossing time, and a lower limit to the event amplitude. Attempts have been made on both of these fronts to interpret the HME in terms of the size of the magnified continuum region (eg. Wambsganss, Paczynski \& Schneider (1990); Rauch \& Blandford 1991; Jaroszynski, Wambsganss \& Paczynski 1992; Webster et al. 1991).  

The amplitude of an HME is a function of source size - smaller sources produce events of larger amplitude. However, the event amplitude is also dependent on the width or strength of a caustic, a quantity that is the constant of proportionality in the near caustic approximation of Chang \& Refsdal (1979). Witt (1990) termed this constant the flux factor and found that it takes on a range of values in the complex caustic structures that are produced by high optical depth microlensing models. The mean value of the flux factor depends on the mass function. Thus the event amplitudes have a distribution that is model dependent. Wambsganss, Paczynski \& Schneider (1990) found that a microlensing model with a mean microlens mass of 0.225$M_{\odot}$ reproduces the observed luminosity variation when the source has a radius smaller than $\sim 2\times 10^{15}cm$. Several attempts have been made to discuss the observed HME amplitude in terms of physical models for the continuum region (eg. Rauch \& Blandford 1991; Jaroszynski, Wambsganss \& Paczynski 1992). The results of these calculations remain controversial.
In addition, Lewis et al. (1998) employ the observed difference in spectra obtained at two different epochs to infer a continuum region that is $\la 10^{15}\,cm$ (based on a blackbody accretion disk model).

A different approach is to use the observed event duration in combination with the transverse velocity to obtain an estimate of source size. Approximate values have been obtained assuming a galactic transverse velocity of $v_{t}\sim 600\,km\,sec^{-1}$ (providing the caustic velocity with respect to the source), and a perpendicular caustic crossing (eg. Irwin et al 1989; Webster et al. 1991). These analyses have yielded values for the continuum source size that are consistent with those obtained through consideration of the event amplitude.

In this work we present an analysis of the source size required to re-produce the observed event duration from model light-curves. This work follows results for the probability of the galactic transverse velocity that we have obtained previously (Wyithe, Webster \& Turner 1999b,c, hereafter WWTb, WWTc). In this analysis we consider the effect on the caustic crossing time of a stellar velocity dispersion and the orientation of caustics with respect to the source trajectory.

\section{The Microlensing model}
\label{model}
This work is based on an extensive suite of numerical microlensing simulations presented previously (WWTc); their salient features are discussed below.
We adopt the optical depth and shear ($\kappa$ and $\gamma$) parameters for the lensing galaxy of Q2237+0305 calculated by Schmidt, Webster \& Lewis (1998), and limit our attention to microlensing models in which all the point masses have identical mass. Four classes of models are produced, combining 0\% and 50\% smoothly distributed matter with source trajectory directions parallel to the A$-$B ($\gamma_{A},\gamma_{B}<0$) and C$-$D ($\gamma_{A},\gamma_{B}>0$) axes. We simulate mock data sets using the sampling rate and period of the available monitoring data (Irwin et al. 1989; Corrigan et al. 1991; $\O$stensen et al. 1995). Two assumptions are made for the photometric error: $\sigma_{SE}$=0.01 mag ($\sigma_{LE}$=0.02 mag) in images A/B and $\sigma_{SE}$=0.02 mag ($\sigma_{LE}$=0.04) in images C/D. We assume that microlensing is produced through the combination of a galactic transverse velocity with an isotropic Gaussian velocity dispersion ($\sigma_{*}\sim 165\,km\,sec^{-1}$ from Schmidt, Webster \& Lewis 1998). Where required, a cosmology having $\Omega=1$ with $H_{0}=75\,km\,sec^{-1}$ is assumed.

This paper builds on work presented in WWTc. Discussions of the microlensing models, the effective transverse velocity ($V_{eff}$), and the method for the determination of its probability $P_v(V_{gal}<V_{eff}|\langle m\rangle)$ where $\langle m\rangle$ is the mean microlens mass can be found in that paper. The galactic transverse velocity is obtained from the effective transverse velocity by comparing the microlensing rates using histograms of light curve derivatives (see WWTb).

\begin{figure}
\vspace*{75mm}
\includegraphics{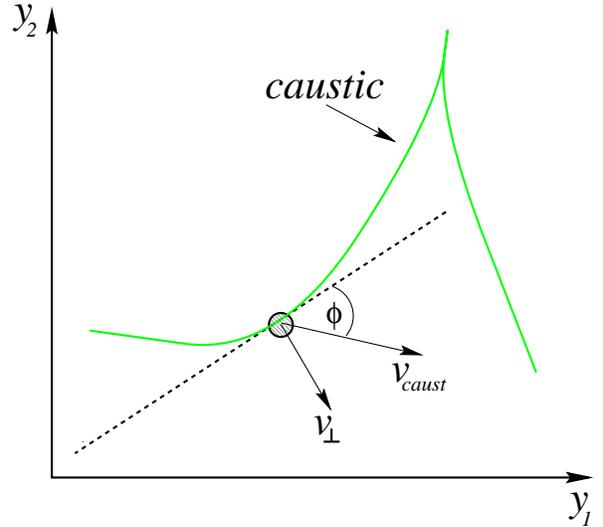}
\caption{\label{cross_geom}A schematic of the  geometry of a caustic crossing a source.}
\end{figure}

\begin{figure*}
\vspace*{65mm}
\includegraphics{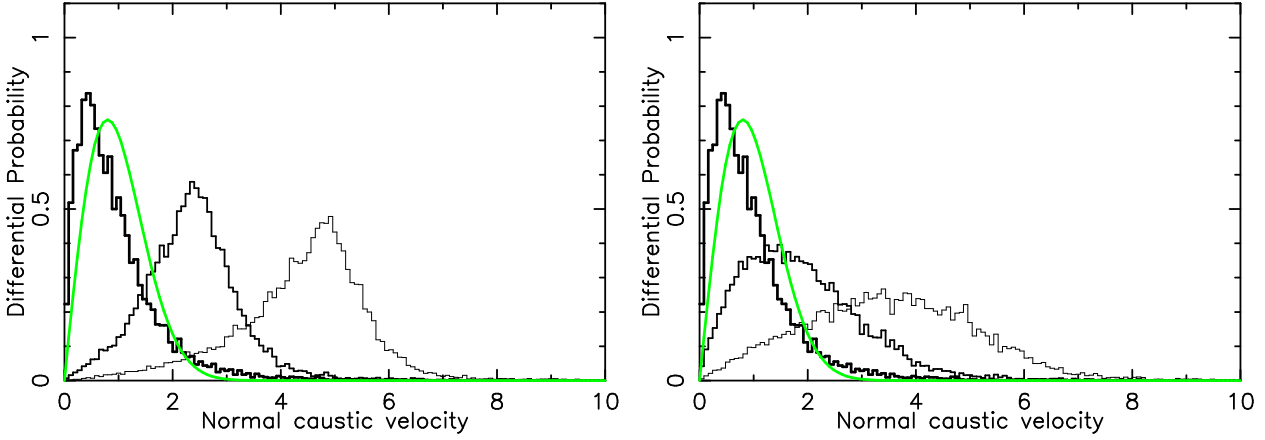}
\caption{\label{caust_prob}The probability distribution functions of normal caustic velocities. The thin, medium and thick lines correspond to transverse velocities of $v_{t}=0,2.5$ and $5\times$ the average projected speed of the stellar velocity dispersion respectively. The thin light line shows the Gaussian distribution of stellar speeds. Left: The applied shear is positive $(\gamma>0)$. Right: The applied shear is negative $(\gamma<0)$.} 
\end{figure*}

\section{Source Size}
\subsection{Source Size determination}

The HME in image A observed during late 1988 is described by 5 observations (Irwin et al. 1989; Corrigan et al. 1991) that constrain a peak in the R-band light-curve. We assume that the length of the HME is twice the estimated rise time of 26 days (Corrigan et al. 1991). The likelihood for the true R-band continuum source size is determined by comparing the observed event duration to distributions of model event durations. The calculations of these distributions include consideration of the probability for effective transverse velocity (WWTc), the normal caustic velocities that arise from the proper motion of stars, and caustic orientations.

The velocity of the caustic network resulting from the stream motion of a starfield was calculated by Schramm et al. (1993). This result was generalised by Kundic, Witt \& Chang (1993) to compute the velocity of individual caustic points due to the proper motion. We use the expressions obtained to find the velocities of caustic points in the $y_{1}$ and $y_{2}$ directions. The vector addition of these components is in general in a direction other than normal to the caustic at that point. Therefore we also calculate the caustic gradient. The method of calculating both the caustic location and gradient are described below.

 We implement the contouring algorithm described by Lewis et al. (1993). The image curves are followed in discrete steps in one direction. When an image line crosses a critical curve, the image magnification changes sign due to the opposite parity of critical images on either side of the critical curve. The location of the critical curve is then found by searching for the image position that produces a lens mapping Jacobian determinant of zero, and used to calculate the caustic velocity. The gradient of the caustic tangent is found by first computing the location of an image position a short distance along the critical curve in both directions. The gradient of the caustic tangent is computed from the corresponding caustic points.

In each of the models considered we found the distributions of caustic velocities and gradients by searching for caustics along an equal length of source track in two orthogonal directions. The source tracks were each $10\eta_{o}$ (corresponding to $\sim 70$ years for a transverse velocity of $600\,km\,sec^{-1}$) in length, where $\eta_o$ is the Einstein radius of a $1M_{\odot}$ star. We computed the properties of the caustics that intersect with 500 such source tracks in directions parallel ($\gamma>0$) and perpendicular ($\gamma<0$ ) to the shear. A separate model starfield was used for each track. This is equivalent to finding the caustic locations on a $500\times 500$ line grid covering a $10\eta_{o} \times 10\eta_{o}$ square of source plane, although the use of 1000 starfields makes the sample more statistically representative.

 Fig.~\ref{cross_geom} is a schematic representation of a caustic crossing, all velocities in this figure are source plane velocities. The caustic velocity $\vec{v}_{caust}$ is the vector addition of the contribution to caustic velocity that results from stellar proper motion $(v_{prop\,1},v_{prop\,2}$), and the transverse velocity $(v_{tran\,1},v_{tran\,2})$,
where
\begin{equation}
(v_{tran\,1},v_{tran\,2})=\vec{v}_{tran}=\frac{D_S}{D_D}\vec{V}_{tran}. 
\end{equation}
$V_{tran}$ is the galactic transverse velocity relative to the observer source line of sight, and $D_S$ and $D_D$ are the Angular diameter distances of the source and lens respectively.
The normal caustic velocity $\vec{v}_{\perp}$ is computed from $\vec{v}_{caust}$ and the relative orientation between the tangent to the caustic and the caustic velocity vector. The magnitude of the normal caustic velocity is:
\begin{equation}
v_{\perp}=|\vec{v}_{\perp}|=|\vec{v}_{caust}|\sin(\phi),
\end{equation}
where $\phi:0\leq\phi\leq\frac{\pi}{2}$ is the angle between the caustic tangent and velocity vector. A caustic with a velocity vector that makes a small angle with the caustic tangent sweeps out a smaller area per unit length per unit velocity than one having a velocity normal to its tangent. Therefore, calculations of the distribution of normal caustic velocities at a stationary source position, include an additional weighting of $\sin(\phi)$.

\begin{figure}
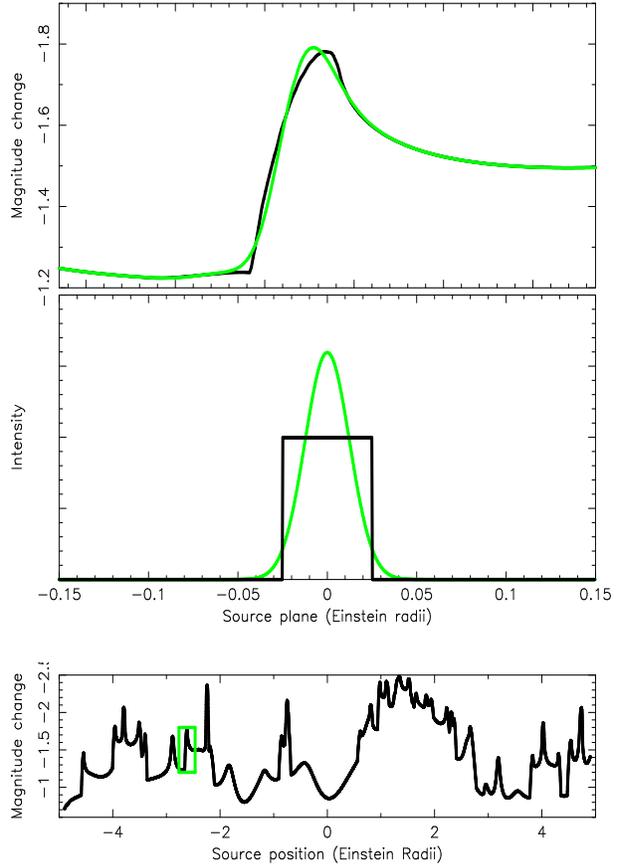

\vspace*{120mm}
\includegraphics{fig3a.epsi}
\includegraphics{fig3b.epsi}
\includegraphics{fig3c.epsi}
\caption{\label{cross} Bottom: A sample light-curve for image A ($\gamma_{A}=+0.4$) of Q2237+0305. One half of a double peaked event is highlighted by a box. Centre: A top-hat intensity profile (dark line), and a Gaussian intensity profile (light line). Top: A plot of the the highlighted HME showing the details of the event shapes that result from the two source intensity profiles.}
\end{figure}

\begin{table*}
\caption{\label{tab_size} Table showing the upper and lower 99\%, 95\%, and 90\% limits as well as the mean of the distributions $p_{s}(S_{s}\,|\,0.1M_{\odot})$ and $p_{s}(S_{s}\,|\,1.0M_{\odot})$. The values are given for the two directions assumed for the transverse velocity as well as for simulations that contain 0\% and 50\% smoothly distributed matter. The limits shown correspond to the cases where the error was $\sigma_{SE}=0.01$ mags in images A/B, $\sigma_{SE}=0.02$ mags in images C/D, and where the error was  $\sigma_{LE}=0.02$ mags in images A/B, $\sigma_{LE}=0.04$ mags in images C/D (in parentheses).}
0\% Smooth Matter
\begin{tabular}{|c|c|c|c|c|c|c|c|c|c|}
\hline
Mean          &Trajectory                &$S_{low}(99\%)$   &$S_{low}(95\%)$   &$S_{low}(90\%)$   &    Mean          &$S_{up}(90\%)$    &$S_{up}(95\%)$    &$S_{up}(99\%)$    \\
Mass          &Direction                 &($\times 10^{14}cm$)&($\times 10^{14}cm$)&($\times 10^{14}cm$)&($\times 10^{14}cm$)&($\times 10^{14}cm$)&($\times 10^{14}cm$)&($\times 10^{14}cm$) \\ \hline\hline
0.1$M_{\odot}$&$\gamma_{A},\gamma_{B}>0$ & 0.244 (0.238)       & 0.637 (0.617)       & 1.01 (0.972)   & 6.70 (6.52)       & 12.1 (11.9)       & 16.6 (16.2)       & 29.2 (29.0)        \\
1.0$M_{\odot}$&$\gamma_{A},\gamma_{B}>0$ & 0.394 (0.280 )      & 1.10  (0.746)       & 1.80 (1.19)    & 12.7 (8.48)       & 24.1 (16.0)       & 31.3 (21.7)       & 49.2 (36.6)        \\
0.1$M_{\odot}$&$\gamma_{C},\gamma_{D}>0$ & 0.240 (0.237)       & 0.622 (0.614)      & 0.982 (0.968)  & 6.56 (6.50)       & 11.9 (11.9)      & 16.3 (16.2)      & 29.1 (29.0)                \\
1.0$M_{\odot}$&$\gamma_{C},\gamma_{D}>0$ & 0.370 (0.272)       & 1.03  (0.723)       & 1.67 (1.16)    & 11.9 (8.11)       & 22.6 (15.2)       & 29.2 (20.5)      & 46.0 (35.1)        \\ \hline

\end{tabular}
\vspace{7 mm}

50\% Smooth Matter
\begin{tabular}{|c|c|c|c|c|c|c|c|c|c|}
\hline
Mean          &Trajectory                &$S_{low}(99\%)$   &$S_{low}(95\%)$   &$S_{low}(90\%)$   &    Mean          &$S_{up}(90\%)$    &$S_{up}(95\%)$    &$S_{up}(99\%)$    \\
Mass          &Direction                 &($\times 10^{14}cm$)&($\times 10^{14}cm$)&($\times 10^{14}cm$)&($\times 10^{14}cm$)&($\times 10^{14}cm$)&($\times 10^{14}cm$)&($\times 10^{14}cm$) \\ \hline\hline
0.1$M_{\odot}$&$\gamma_{A},\gamma_{B}>0$ & 0.207 (0.187)       & 0.556 (0.493)      & 0.895 (0.795)  & 5.95 (5.31)       & 11.1 (9.75)       & 14.6 (13.3)       & 24.0 (23.2)        \\
1.0$M_{\odot}$&$\gamma_{A},\gamma_{B}>0$ & 0.487 (0.274)       & 1.45  (0.772)       & 2.48 (1.26)    & 16.4 (10.1)       & 30.1 (19.4)       & 38.1 (25.7)       & 58.4 (40.4)        \\ 
0.1$M_{\odot}$&$\gamma_{C},\gamma_{D}>0$ & 0.206 (0.188)       & 0.553 (0.495)       & 0.888 (0.796)  & 5.73 (5.27)       & 10.5 (9.57)       & 13.1 (13.2)       & 23.3 (23.0)        \\
1.0$M_{\odot}$&$\gamma_{C},\gamma_{D}>0$ & 0.468 (0.272)       & 1.36  (0.760)       & 2.26 (1.23)    & 14.5 (9.25)       & 26.8 (17.7)       & 33.7 (23.2)       & 49.3 3(6.7)        \\ \hline
\end{tabular}
\end{table*}

Figure \ref{caust_prob} shows probability distributions of normal caustic velocities $p_c(v_{\perp}|\langle m\rangle, v_{tran})$ where $v_{tran}=|\vec{v}_{tran}|$. The left and right hand diagrams correspond to the cases of positive and negative galactic shear respectively. Distributions are shown in each case for the normal caustic velocities due to the stellar proper motions alone (thick line), and in combination with a transverse velocity in the $y_{1}$ direction of 2.5 (medium line) and 5.0 (thin line). Also plotted in Fig.~\ref{caust_prob} is the distribution of stellar speeds (light line). All velocities are in units of the projected mean speed of the isotropic Gaussian stellar velocity dispersion.
 These distributions demonstrate the dependence of the normal caustic velocity on the direction of transverse velocity; a transverse velocity along the direction of caustic clustering ($\gamma<0$) results in lower normal caustic velocities because the tangent to the caustic is more likely to lie along the source trajectory. Note however that the distribution must be independent of the shear direction if the transverse velocity is zero. 

We assume a circular projected source profile so that the event timescale is not dependent on the source orientation. We define the source diameter $S_{s}$. For a small source ($S_{s}\ll\eta_{o}$) the crossing time of the caustic, and therefore the event timescale is
\begin{equation}
\label{size}
\Delta T\sim S_{s}/v_{\perp}.
\end{equation}
 Thus, $S_{s}$ can be estimated by combining a measured timescale with statistical information about caustic orientations and velocities. Eqn \ref{size} demonstrates a systematic uncertainty in our determination of source size that is proportional to the uncertainty in the estimate of the event duration.

Eqn \ref{size} defines our measure of source size, however its interpretation is dependent on the intensity profile of the source. The lower panel in Fig.~\ref{cross} shows an example of an extended source light-curve for image A ($\gamma_{A}=+0.4$) computed using the method described in Wyithe \& Webster (1999). One half of a double peaked event is highlighted by a box and reproduced in the top panel of Fig.~\ref{cross} for two source intensity profiles: a Gaussian profile (light line) and a top-hat profile (dark line). Cross-sections of the intensity profiles are shown in the central panel of Fig.~\ref{cross} and comparison of these with the corresponding HMEs demonstrates that $S_{s}$ is approximately the source diameter for a top-hat source profile, but twice the half width $S_{s}$ for a Gaussian profile.

Probability distributions for normal caustic velocity $p_{c}(v_{\perp}\,|\,\langle m\rangle,v_{tran})$ are computed assuming various values for the galactic transverse velocity. By differentiating the cumulative distribution $P_v(V<V_{tran}|\langle m\rangle)$ (following a correction for the contribution of stellar proper motion to the value of effective transverse velocity (WWTb)), we also find the probability distribution for galactic transverse velocity, 
\begin{equation}
p_{v}(v_{tran}=\frac{D_S}{D_D}V_{tran}\,|\,\langle m\rangle)=\frac{D_D}{D_S}p_v(V_{tran}|\langle m\rangle).
\end{equation}
The probability distribution for source size $S_s$ is

\begin{eqnarray}
&\hspace{-42mm}p_{s}(S_{s}=v_{\perp}\Delta T\,|\,\langle m\rangle)= \nonumber \\
&\hspace{5mm}\frac{1}{\Delta t}\,\int p_{c}(v_{\perp}\,|\,\langle m\rangle,v_{tran})\,p_{v}(v_{tran}\,|\,\langle m\rangle)\,dv_{tran} .
\end{eqnarray}

\subsection{\label{source_size}Source size limits}

\begin{figure*}
\vspace*{130mm}
\includegraphics{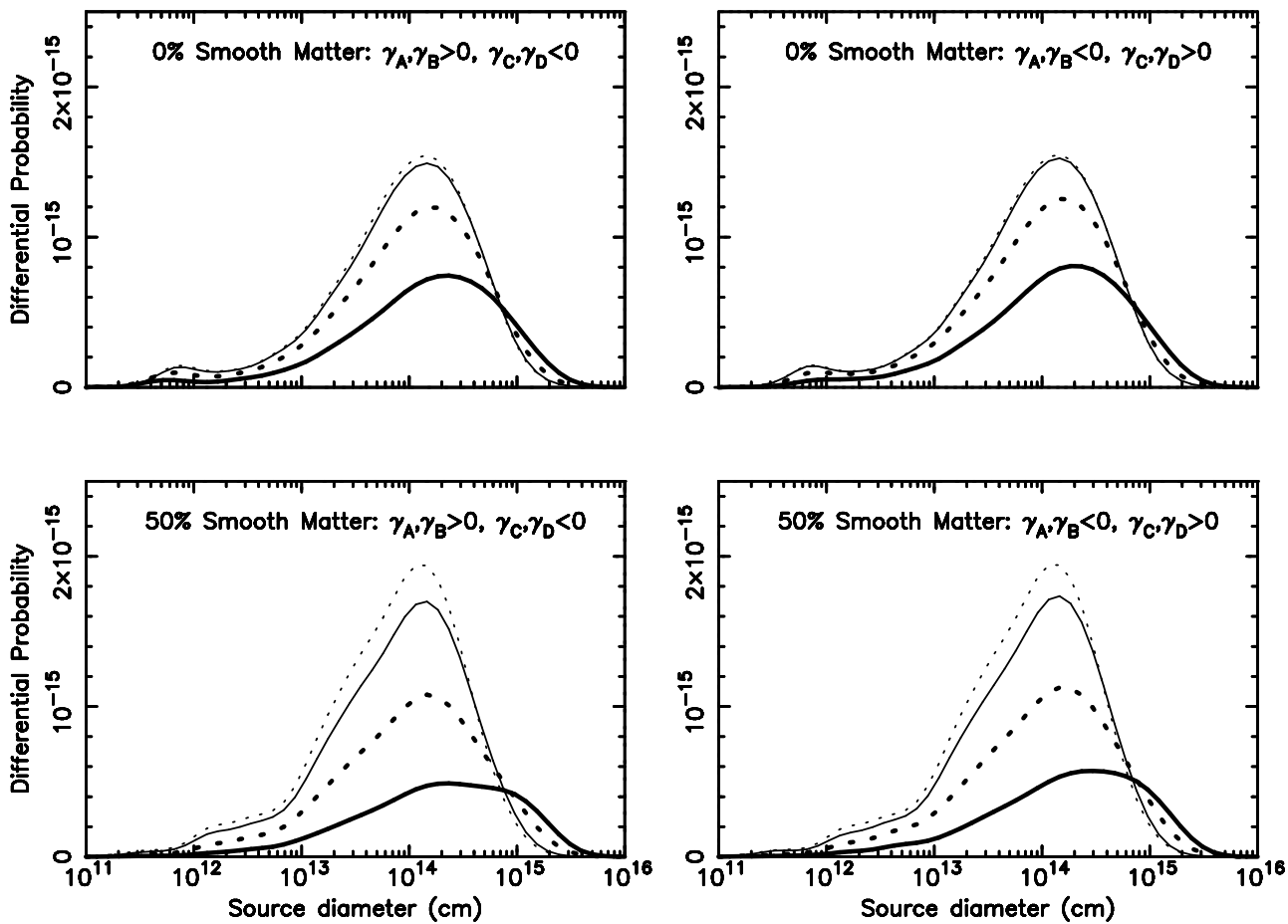}
\caption{\label{diff}The differential distributions of source size $p_s(S_{s}\,|\,0.1M_{\odot})$ (thin lines) and $p_s(S_{s}\,|\,1.0M_{\odot})$ (thick lines). The solid and dotted curves represent the resulting functions when the photometric error was assumed to be $\sigma_{SE}$=0.01 mag in images A/B and $\sigma_{SE}$=0.02 in images C/D, and $\sigma_{LE}$=0.02 mag in images A/B and $\sigma_{LE}$=0.04 in images C/D.}
\end{figure*}

\begin{figure*}
\vspace*{130mm}
\includegraphics{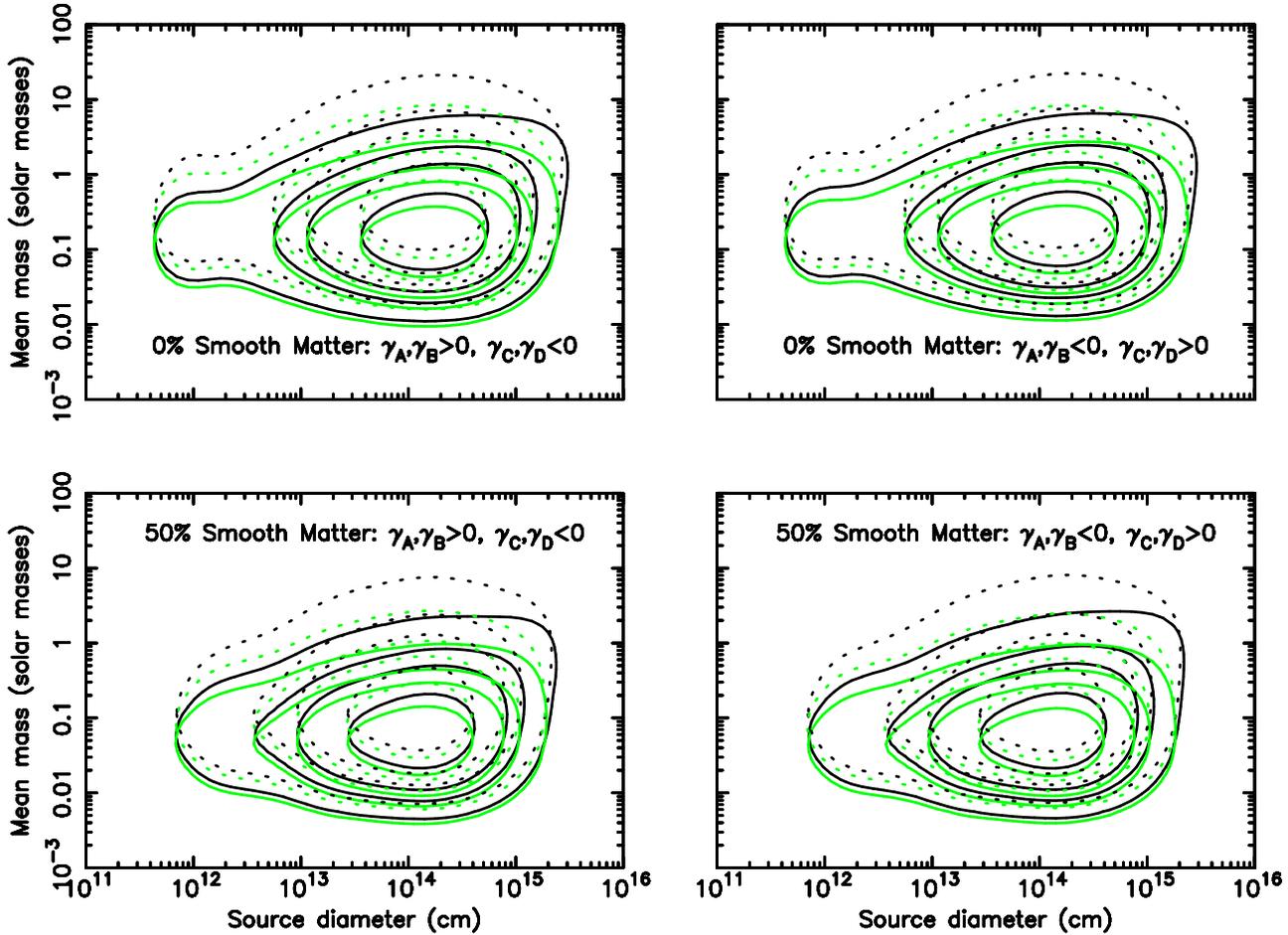}
\caption{\label{contour}Contours of percentage peak height of the function $p_{m,s}(S_{s},\langle m\rangle)$. The contours shown are the 3.6\%, 14\%, 26\% and 61\% levels. The solid and dotted curves represent the resulting functions when the photometric error was assumed to be $\sigma_{SE}$=0.01 mag in images A/B and $\sigma_{SE}$=0.02 in images C/D, and $\sigma_{LE}$=0.02 mag in images A/B and $\sigma_{LE}$=0.04 in images C/D. The light and dark lines correspond to the assumptions of logarithmic and flat priors for transverse velocity referred to in the text.}
\end{figure*}

Our results are presented in Tab.~\ref{tab_size}. Values are given for the upper and lower 99\%, 95\%, and 90\% limits as well as the mean of the distributions $p_{s}(S_{s}\,|\,0.1M_{\odot})$ and $p_{s}(S_{s}\,|\,1.0M_{\odot})$ for models having various combinations of assumptions for the transverse velocity directions and smooth matter fraction. Limits are shown assuming the small (SE) and large (LE, in parentheses) errors defined in Sec.~\ref{model}. 
The differential distributions of source size $p_s(S_{s}\,|\,0.1M_{\odot})$ (thin lines) and $p_s(S_{s}\,|\,1.0M_{\odot})$ (thick lines) are plotted in Fig.~\ref{diff}. The solid and dotted lines correspond to the cases of small and large errors respectively. We find that the source size lies between the values of $2\times 10^{13}cm$ and $6\times 10^{15}cm$ (98\% level).

While the lower source size limit is not very sensitive to the microlens mass assumed, the upper limit is higher for larger assumptions of mean microlens mass. The trend is demonstrated in Fig.~\ref{diff}, and is due to the limit of effective transverse velocity being proportional to $\sqrt{\langle m\rangle}$. Lower limits for source size come from the lowest velocity caustics, which are produced by the stellar proper motions alone, hence the insensitivity to the microlens mass. However the upper limits result from the fastest moving caustics, and so depend on the microlens mass dependent upper limits to transverse velocity. The effect is magnified if a component of smooth matter is assumed. Due to the more sparsely distributed caustic network produced by such a model, a larger transverse velocity is measured (WWTc). Also the contribution of stellar proper motions is reduced. These factors result in an increased sensitivity of source size estimates to microlens mass if the model smooth matter content is higher. The source size distributions are highly asymmetric. However, due to the stability of the lower end of the distributions, the mode is consistent, between $\sim 1\times 10^{14}cm$ and $4\times 10^{14}cm$ for all models considered.

It is apparent from Fig.~\ref{diff} that the assumptions for photometric error and the direction of the transverse velocity are important factors, particularly for a larger assumed mean microlens mass. Like the assumptions for microlens mass, assumptions for the error and transverse velocity direction effect the measurement of transverse velocity, and hence source size. Upper limits are more sensitive than lower limits which are set primarily by the stellar velocity dispersion. These effects are also enhanced by the presence of a component of smooth matter. 

The co-dependence of mean microlens mass and source size is now treated more rigorously.

\subsection{Simultaneous limits of mean microlens mass and source size}

In Sec~\ref{source_size} the normalised probability distributions, $p_s(S_{\rmn s} | \langle m \rangle)$, that the continuum source size is between $S_{\rmn s}$ and $S_{\rmn s} + {\rmn d}S_{\rmn s}$ given that the mean microlens mass is $\langle m \rangle$, were obtained. The probability that the source size lies in the above range and the mean microlens mass is between $\langle m \rangle$ and $\langle m \rangle + {\rmn d} \langle m \rangle$ is then given by
\begin{equation}
p_{m,s}(S_{\rmn s}, \langle m \rangle) = p_s(S_{\rmn s} | \langle m \rangle)
\, p_m (\langle m \rangle),
\end{equation}
where $p_m(\langle m \rangle)$ is the probability that the mean microlens mass is between $\langle m \rangle$ and $\langle m \rangle + {\rmn d} \langle m \rangle$. $p_m(\langle m \rangle)$ is a known function (given in WWTc), that is approximately independent of the source size assumed and was determined under the same assumptions for the Bayesian prior (flat and logarithmic) of galactic effective transverse velocity as the function $p_{v}(v_{tran}|\langle m\rangle)$. However unlike $p_{v}(v_{tran}|\langle m\rangle)$, $p_m (\langle m \rangle)$ is dependent on the prior assumed (see WWTc).

Figure~\ref{contour} shows contour plots of the two-dimensional distribution $p_{m,s}(S_s,\langle m\rangle)$ for the assumed source orientations, smooth matter fraction and photometric errors. All possible combinations of the two values of each parameter are included, so that eight models are shown. The contours are at 3.6, 14, 26, 61 per cent of the peak height, and so the extrema of the relevant contours represent $4\sigma$, $3 \sigma$, $2\sigma$, and $1\sigma$ limits on the single variables. The light and dark lines correspond to the assumptions of logarithmic and flat priors for effective transverse velocity. The most likely model is that $\langle m \rangle \simeq 0.05-0.2 M_{\odot}$ (dependent on the model systematics of trajectory direction, smooth matter content and photometric uncertainty) and $S_{\rmn s} \simeq 2\times10^{14}$ cm (approximately independent of model). 
 
The most important benefit of combining the distributions for $S_{\rmn s}$ and $\langle m \rangle$ is that an upper limit can be placed on the mean microlens mass. The limits on $\langle m \rangle$ are subject to both random and systematic uncertainties; the measured mass increases with photometric errors and decreases with the fraction of smooth matter. Assuming that the chosen values for these parameters bracket the likely models, upper and lower limits for $\langle m \rangle$ can be found by using the most extreme values of the contour limits. This yields the 95 per cent result that $0.01 M_{\odot} \la \langle m \rangle \la 1 M_{\odot}$. The upper limit, which agrees with that obtained from simultaneous consideration of probabilities for $\langle m\rangle$ and $V_{tran}$ (WWTc) is lower than that obtained by Lewis \& Irwin (1996), although sufficiently high that it does not conflict with any popular models of galactic halos or bulges. Similarly, the contour maxima yield the result that there is a 95 per cent chance of $5 \times 10^{12}$ cm $\la S_{\rmn s} \la 10^{15}$ cm. At a given confidence level, both the upper and lower limits of source size, and the lower limit of mean microlens mass have smaller values in Fig.~\ref{contour} than those obtained from probabilities for single variables in previous sections and WWTc. This can be attributed to the asymmetry of the distribution in both dimensions. 

\section{Conclusion}
The observed peak in the light-curve for image A of Q2237+0305 in 1988 appears to be one half of a double peaked event that is characteristic of a source having passed inside of a cusp. If this interpretation is correct then the observed peak is the result of the source having been crossed by a single caustic, and the observations provide a good record of the caustic crossing time. The source size is the product of the event length and the normal velocity of the caustic involved. We have placed limits on normal caustic velocities by combining measurements of the galactic transverse velocity with calculations of the distributions of caustic gradients and the component of the caustic velocity that results from stellar proper motions. We find that the most likely size of the magnified continuum region is $S_{s}\sim 2\times 10^{14}\,cm$, and that the upper and lower limits are $S_{s}\sim 6\times 10^{15}\,cm$ and $S_{s}\sim 2\times 10^{13}\,cm$ (99\% level) respectively. Our calculations are quite insensitive to the systematic considerations of trajectory direction, smooth matter content and assumption of Bayesian prior for effective transverse velocity (within the ranges considered). However, our conclusions are sensitive to the microlens mass assumed, as well as to the error assumed for the published photometric magnitudes. This sensitivity arises from the dependence of the determination of effective transverse velocity on these quantities.

Our conclusions are consistent with those of Wambsganss, Paczynski \& Schneider (1990) and Rauch \& Blandford (1991) who found that the event rise-time and peak height require a source smaller than $\sim2\times10^{15}\,cm$ (if stellar mass objects are assumed for the microlenses). The microlensing determinations from Q2237+0305 support length scales of $ct\sim 10^{14}\,t_{hrs}\,cm$ associated with X-ray and optical changes in continuum emission. Also, theoretical calculations of the typical scale-size of a continuum emitting accretion disc orbiting a supermassive black-hole (Rees 1984) suggest a size of $\sim 10^{15}\,cm$.

 Simultaneous consideration of microlens mass and source size provides an upper limit to the mean microlens mass of $1M_{\odot}$ that is unavailable from consideration of the microlensing rate alone. We find that $0.1M_{\odot}<\langle m\rangle<1M_{\odot}$ (95\% level). Unlike the source size, our mass determinations are dependent on the prior assumed for effective transverse velocity (see Wyithe, Webster \& Turner 1999c).

It was noted in (Wyithe, Webster \& Turner 1999b) that the statistical determination of the effective transverse velocity will be aided when monitoring data from a longer period becomes available. The narrowing of the probability for the effective transverse velocity will in turn provide tighter limits on the mean microlens mass and source size. At a known transverse velocity, the component of uncertainty due to the range of event lengths that can be produced by a source will be reduced upon the observation of more HME durations. Thus the extent of the contours in Fig.~\ref{contour} will be reduced as more monitoring data becomes available.

\section*{Acknowledgements}
The authors would like to thank Chris Fluke for helpful discussions. This work was supported in part by NSF grant AST98-02802. JSBW acknowledges the support of an Australian Postgraduate award and a Melbourne University Overseas Research Experience Award.

\label{lastpage}

\end{document}